# Active Magnetoplasmonics with Transparent Conductive Oxide Nanocrystals

Alessio Gabbani,[1,2,3] Claudio Sangregorio,[2,3] Bharat Tandon,[4] Angshuman Nag,[4] Massimo Gurioli,[5] Francesco Pineider[1*]

[1] INSTM and Department of Chemistry and Industrial Chemistry, Università di Pisa, via G. Moruzzi 13, 56124, Pisa Italy
[2] INSTM and Department of Chemistry "U. Schiff", Università degli Studi di Firenze, via della Lastruccia 3, 50019 Sesto Fiorentino (FI), Italy
[3] CNR-ICCOM, Via Madonna del Piano 10, 50019 Sesto Fiorentino (FI), Italy
[4] Department of Chemistry, Indian Institute of Science Education and Research (IISER), Pune 411008, India
[5] Department of Physics and Astronomy, 672 Università degli Studi di Firenze, 50019 Sesto Fiorentino, FI, 673 Italy
*Corresponding author: francesco.pineider@unipi.it

**Abstract**

**Magnetoplasmonics is highly promising to devise active optical elements: modulating the plasmon resonance condition with magnetic field can boost the performance of refractometric sensors and nanophotonic optical devices. Nevertheless, real life applications are hampered by the magnetoplasmonic trilemma: 1) a good plasmonic metal has sharp optical resonances but low magneto-optical response; 2) a magnetic metal has strong magneto-optical response but a very broad plasmonic resonance; 3) mixing the two components degrades the quality of both features. To overcome the trilemma, we use a different class of materials, transparent conductive oxide nanocrystals (NCs) with plasmonic response in the near infrared. Although non-magnetic, they combine a large cyclotron frequency (due to small electron effective mass) with sharp plasmonic resonances. We benchmark the concept with F- and In- doped CdO (FICO) and Sn-doped $In_2O_3$ (ITO) NCs to boost the magneto-optical Faraday rotation and ellipticity, reaching the highest magneto-optical response for a non-magnetic plasmonic material, and exceeding the performance of state-of-the-art ferromagnetic nanostructures. The magnetoplasmonic response of these NCs was rationalized with analytical model based on the excitation of circular magnetoplasmonic modes. Finally, proof of concept experiments demonstrated the superior performance of FICO NCs with respect to current state of the art in magnetoplasmonic refractometric sensing, approaching the sensitivity of leading localized plasmon refractometric methods with the advantage of not requiring complex curve fitting.**

**Keywords** Magnetoplasmonics, Active Plasmonics, Transparent Conductive Oxides Nanocrystals, Magneto-optics, Faraday rotation and ellipticity, Refractometric Sensing

**Main**

Active plasmonics, the modulation of plasmons resonances, is at the forefront of today's research in nano-optics.[1,2] Active control of light at the subwavelength scale can enable ultrahigh performance optical sensors based of refractometry[3–5] or field enhancement,[6,7] as well as tuneable nanophotonic optical components.[8–10] Several elegant approaches have been proposed to achieve active plasmonics, through controlled modification of the refractive index of the medium,[11,12] mechanical deformation of the substrate,[13,14] or via electric field or electrochemical stimuli,[15–17] to name a few. Unfortunately, none of these approaches satisfies the three key requirements (speed, reversibility and ease of implementation) at the same time. Magnetic fields are easy to generate and propagate, and their effects on charge carriers are ultrafast and fully reversible: their use in combination with polarized light to control plasmon resonances (magnetoplasmonics[18–22]) is thus a strong candidate for active plasmonics.

For the design of efficient active magnetoplasmonic elements, two key factors come into play: the magnitude of the magnetic modulation and the quality of the optical resonance, i.e. its line width and extinction coefficient. Using magneto-optical spectroscopic techniques, magnetically-driven modulation of localized surface plasmon resonance (LSPR) was recently observed on different combinations of materials, from pure noble metals[23–29] to ferromagnetic or hybrid noble metal/ferromagnetic nanostructures.[18,30–34] Nevertheless, achieving strong magnetic modulation without degrading the plasmonic properties remains challenging. Indeed, the introduction of a ferromagnetic metal gives larger magnetic modulation, which is proportional to the magnetization of the material at the cost of strong optical losses resulting in broader and weaker resonances. On the other hand, nanostructures made of plasmonic metals (e.g. Au, Ag) are characterized by sharper resonances, but display a weaker magnetic modulation with respect to the former ones. In noble metals nanostructures the modulation is proportional to the cyclotron frequency ($\omega_c = eB/m$),[25] where $e$ and $m$ are the charge and effective mass of electron, respectively, and $B$ is the applied magnetic field (assuming zero magnetization of the material). A crucial parameter to improve the magneto-optical response is thus represented by $m$. For most metals $m$ has a fixed value, close to the free electron mass $m_e$; conversely, in heavily doped semiconductors, carrier parameters (charge, density and mass) can be modulated by doping.[35–40] Among these materials, Transparent Conductive Oxides (TCO), an emerging class of infrared plasmonic materials,[41–43] display reduced carrier effective mass ($m$) with respect to noble metals (down to $0.4$-$0.2 m_e$),[35,36] which makes them very attractive for magneto-plasmonic. Nevertheless, the application of TCO nanostructures in this field has not been explored to date.

Here we demonstrate TCO-based devices for magnetic field-modulated active plasmonics. Colloidal dispersions of TCO nanocrystals (NCs) showed a magnetoplasmonic modulation which is unprecedented in non-magnetic plasmonic materials and is strongly competitive with state-of-the-art magnetic-plasmonic architectures, while retaining superior resonance quality factors. These features prompt improved refractometric sensing platforms with performance indexes challenging both optical and magneto-optical approaches. Monitoring the Faraday ellipticity of F- and In- doped CdO (FICO) NCs at fixed wavelength affords a sensitivity of 1.24 deg per refractive index unit, potentially detecting refractive index changes down to $3\times10^{-6}$, without requiring a fitting approach. To the best of our knowledge this is the first report of the TCO NCs for magnetoplasmonics and the first time that their magnetoplasmonic performances for refractometric sensing have been estimated.

Among TCOs, Sn-doped $In_2O_3$ (ITO) and FICO NCs have been chosen: the first one is the most established TCO and has a superior magneto-optical response compared to noble metal NCs;[44] the second one has a significantly reduced LSPR linewidth, thanks to cooperative anion-cation co-doping,[45,46] which is expected to further improve the magneto-optical response. Both semiconductors have a relatively small electron effective mass ($\approx 0.3$-$0.4 m_e$).[44,47]

Quasi-spherical ITO and FICO NCs with plasmonic resonance in the same wavelength range were synthesized by colloidal chemistry approaches (see methods). Transmission electron microscopy (TEM) morphological and structural analysis reveal a mean size of $9.0 \pm 1.7$ nm and $15.0 \pm 2.3$ nm for ITO and FICO NCs respectively ((**Figure 1** a-b and **Figure S1**). The X-ray diffraction peaks (**Figure 1** c) show negligible shifts with respect to those of undoped $In_2O_3$ and CdO respectively, without secondary crystalline phases. The aliovalent dopant incorporation was confirmed by Inductively Coupled Plasma Atomic Emission Spectroscopy (ICP-AES), revealing a content of 10% of Sn and 26% of In for ITO and FICO NCs respectively. The optical extinction spectra collected in a $CCl_4$ dispersion (**Figure 2** a-b) revealed the presence of an electric dipole plasmonic resonance (with negligible scattering contribution due to the reduced NCs size compared to the wavelength of the impinging light) at near infrared wavelengths with peaks at 1825 nm (0.679 eV) and 1883 nm (0.659 eV) for ITO and FICO NCs, respectively. The LSPR of the two NCs display significantly different full-width-half-maximum (FWHM), which are 0.19 eV and 0.10 eV respectively for ITO and FICO NCs. This difference is justified by the different electron scattering with dopant impurities in the two semiconductors.[45–48] No effect of the NCs size and polydispersity on the LSPR linewidth is predicted by Mie theory given the small diameter of the NCs compared to incident wavelength.[49,50] FICO NCs display reduced LSPR damping, providing sharper LSPR peaks than ITO and a higher quality factor of the resonance.

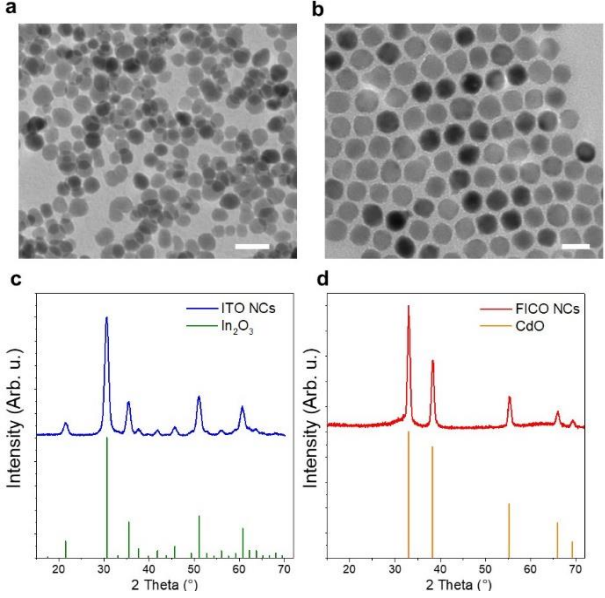

**Figure 1**: **Morphology and crystal structure.** Representative TEM images of ITO (a) and FICO NCs (b). Scale bar is 20 nm. Powder X-ray diffraction pattern collected for ITO (c) and FICO NCs (d). The reference pattern of $In_2O_3$ (PDF 06-0416) and CdO (PDF 05-0640) are shown for comparison.

The full magneto-optical response was investigated at room temperature by measuring the Faraday rotation ($\theta_F$) and ellipticity ($\varepsilon_F$) spectra of the NCs dispersed in $CCl_4$ (Figure 2 c,d). $\varepsilon_F$ feature a dispersive shape, crossing the zero at the LSPR energy ($E_0$). This is consistent with a helicity-dependent opposite shift of the resonance condition with respect to $E_0$, which is driven by the applied magnetic field. On the other hand, $\theta_F$ displays a positive maximum at the LSPR energy, with two weaker negative peaks, one at higher energy and the other at lower energy with respect to $E_0$. Both magneto-optical signals display a linear field-dependence, as expected for non-magnetic plasmonic nanostructures.

The optical response of plasmonic nanospheres can be calculated from the quasi-static polarizability of a sphere ($\alpha$).[49,50] The effect of magnetic field on LSPR can be rationalized in terms of field-driven modification of the NC polarizability (Equation S 4 and ref. [51]), resulting in the splitting of the otherwise degenerate circular plasmonic modes, as already reported for noble metal nanostructures.[25,28,51] An analytical expression for the helicity-dependent differential polarizability is thus obtained (under an applied magnetic field) ($\Delta\alpha_B(\omega) = \alpha_{RCP}(\omega) - \alpha_{LCP}(\omega)$). From this expression, the Faraday ellipticity and rotation can be calculated through Equations (1) and (2) (see section 2 of the supporting information for more details).

(1) $\varepsilon_F(deg) = ln10 \cdot \frac{\Delta A}{4} \cdot \frac{180}{\pi} = ln10 \cdot \frac{180}{4\pi} k\sqrt{\varepsilon_m} I[\Delta\alpha_B(\omega)]$

(2) $\vartheta_F(deg) = ln10 \cdot \frac{180}{4\pi} \cdot k\sqrt{\varepsilon_m} R[\Delta\alpha_B(\omega)]$,

where $k$ is the wavevector of light, $\varepsilon_m$ is the solvent permittivity, and ($ln10 \cdot \frac{180}{4\pi}$) is the conversion factor of the signal from MCD ($\Delta A$ units) into ellipticity angle (degrees).[52]

The dielectric function of the NCs can be expressed in terms of the carrier parameters $N$, $m$ and $\gamma$, according to the Drude model (see section S2 of the supporting information for details),[48,50] and inserted into the quasi-static field-dependent polarizability (Equation S 4). Thus, using the analytical equations, we extract the fundamental parameters of the charge carriers involved in LSPR, i.e. carrier density $N$, mass $m$, and damping parameter $\gamma$, through the simultaneous fitting of the normalized ellipticity and extinction spectra (more details are provided in section S2 of the supporting information). The determination of the Drude parameters is quite critical, as they are the main factors determining the optical and magneto-optical response. The ratio $N/m$ controls the LSPR position, whereas $m$ is inversely proportional to the magnetic modulation of LSPR. Using the obtained Drude parameters (reported in **Table S1**), the Faraday ellipticity and rotation can be calculated, obtaining an excellent agreement with the experimental data (**Figure 2** e-f). Carrier densities of 7.13 and 9.24x10$^{20}$ cm$^{-3}$ were obtained for ITO and FICO NCs, roughly two order of magnitudes lower then Au. The comparison with Au NCs (**Table S1**) also shows that the reduced values of effective mass (0.27$m_e$ and 0.31$m_e$ for ITO and FICO NCs respectively) are the main cause for the significantly boosted magneto-optical signal (20-30 fold enhancement) in transparent conductive oxide NCs. An important role is also played by the broadening of the LSPR peak (in first approximation related to $\gamma$): this is clearly shown by the stronger magneto-optical signal of FICO with respect to ITO NCs. In fact, the two materials display comparable effective mass but different LSPR linewidth. These findings highlight the two fundamental requirements to boost magnetoplasmonic modulation in non-magnetic plasmonic NCs: high cyclotron frequency (achievable in materials with low electron effective mass), and reduced LSPR linewidth. A magnetoplasmonic quality factor can thus be defined as the ratio between cyclotron frequency and LSPR linewidth ($\omega_c/\gamma$), reported in Table S1 for the TCO NCs of this study compared to Au colloidal NCs.

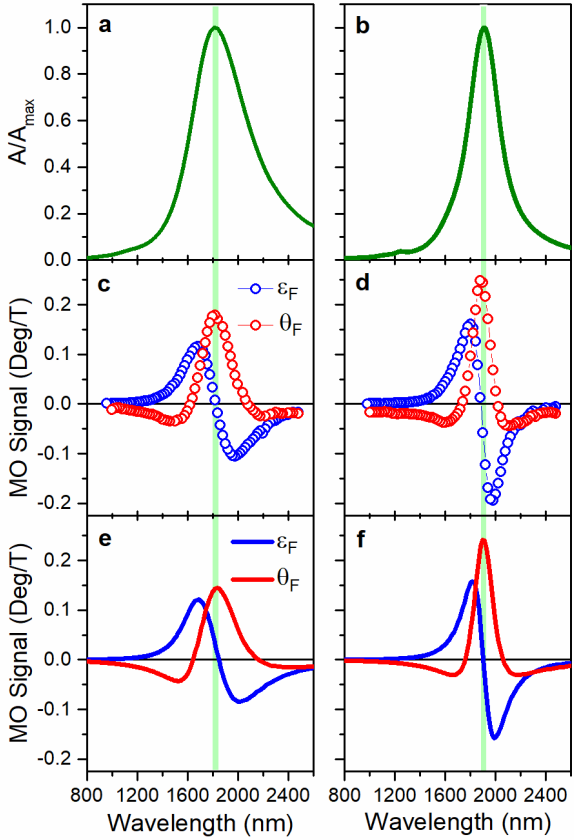

**Figure 2: Magneto-optical properties.** Extinction spectra of ITO (a) and FICO (b) NCs dispersed in CCl$_4$, normalized for the extinction maximum (A/A$_{max}$); experimental (c-d) and calculated (e-f) Faraday ellipticity ($\varepsilon_F$, blue) and rotation ($\theta_F$, red) of the same samples. Both the experimental and calculated magneto-optical responses are normalized to the extinction maximum for the sake of comparison and reported in deg for unit (Tesla) of applied field. The measurements were carried out at room temperature. The vertical pale green line indicates the LSPR resonance wavelength.

To our knowledge, we obtained the highest magneto-optical response for a non-magnetic plasmonic nanostructure, reaching values close to some ferromagnetic or hybrid noble metal-ferromagnetic nanostructures at room temperature and relatively low magnetic fields (1 Tesla). Indeed, signals of 0.15-0.28 deg (for 1 Tesla of applied field) were obtained for FICO NCs, which are of close to those reported for Ni nanodisks (≈0.5 deg at 0.4 Tesla)[31] and larger than Au/Co/Au sandwich nanodisks structures (0.1-0.2 deg at 1 Tesla)[33]. FICO NCs can benefit from their sharper features in the magneto-optical spectrum: this is interesting for sensing, because a steeply sloping signal at the resonance condition will boost the sensitivity of refractometric sensors.

Inspired by the above considerations, we performed a proof of concept refractometric sensing experiment by dispersing the NCs in three solvents which are transparent and have different refractive index (RI) in the spectral range of interest (**Figure S5**): CCl$_4$ (RI=1.4477), C$_2$Cl$_4$ (RI=1.4895) and CS$_2$ (RI=1.5866). Extinction, Faraday rotation and ellipticity spectra of the NCs dispersions were measured for both FICO and ITO NCs. The RI sensitivity obtained in extinction spectroscopy (**Figure S6**) by tracking the shift of the LSPR wavelength is larger than Au nanospheres and comparable to ferromagnetic magnetoplasmonic systems (**Table S2**). The positive peak of the Faraday Rotation spectrum (**Figure 3** a,b) and the zero crossing of the ellipticity (**Figure 3** c,d) also red shift with the increasing refractive index of the medium. Moreover, if we monitor the change in intensity of the magneto-optical signal at a fixed wavelength, a drastic variation occurs with RI, as the resonance is shifted and the slope of the signal ($\delta\varepsilon_F/\delta\lambda$ and $\delta\theta_F/\delta\lambda$) is very high due to the large cyclotron frequency and the reduced LSPR line width. In **Figure 3** e,f, the variation of ellipticity and rotation as a function of RI is reported at a fixed wavelength for an applied field of 1.4 Tesla. The RI sensitivity is calculated as the slope of the linear fit (i.e. $\delta\varepsilon_F/\delta RI$ and $\delta\theta_F/\delta RI$) and compared with the values achieved in the literature for the state of the art magnetoplasmonic nanostructures (summarized in **Table S3**). ITO and FICO NCs showed a sensitivity of 0.29 and 1.24 deg/RIU respectively using the ellipticity signal, whereas values of 0.22 and 1.12 deg/RIU are obtained using the rotation signal. The best performance is displayed by the ellipticity signal of FICO NCs, reaching a RI sensitivity which is 40-fold enhanced with respect to our previous work on Au NPs,[25] 2.5-fold enhanced with respect to Ni nanodisks,[31] and superior to Au/SiO$_2$/Ni multi-layered nanodisks arranged in a periodic array exploiting surface lattice resonance to boost the sensitivity (**Figure 3** e,f and **Table S3**).[53]

The enhanced RI sensitivity achieved with FICO NCs is even more remarkable if we consider that it has been obtained with simple colloidal NCs, without requiring a complex multi-component architecture. On the other hand, ITO NCs achieved lower sensitivity, due to the reduced slope of the magneto-optical signal at the resonance condition. This can be ascribed to the decreased LSPR quality factor of ITO due to the higher electron scattering by ionic impurities, pointing out the importance of sharp resonances for this sensing approach.

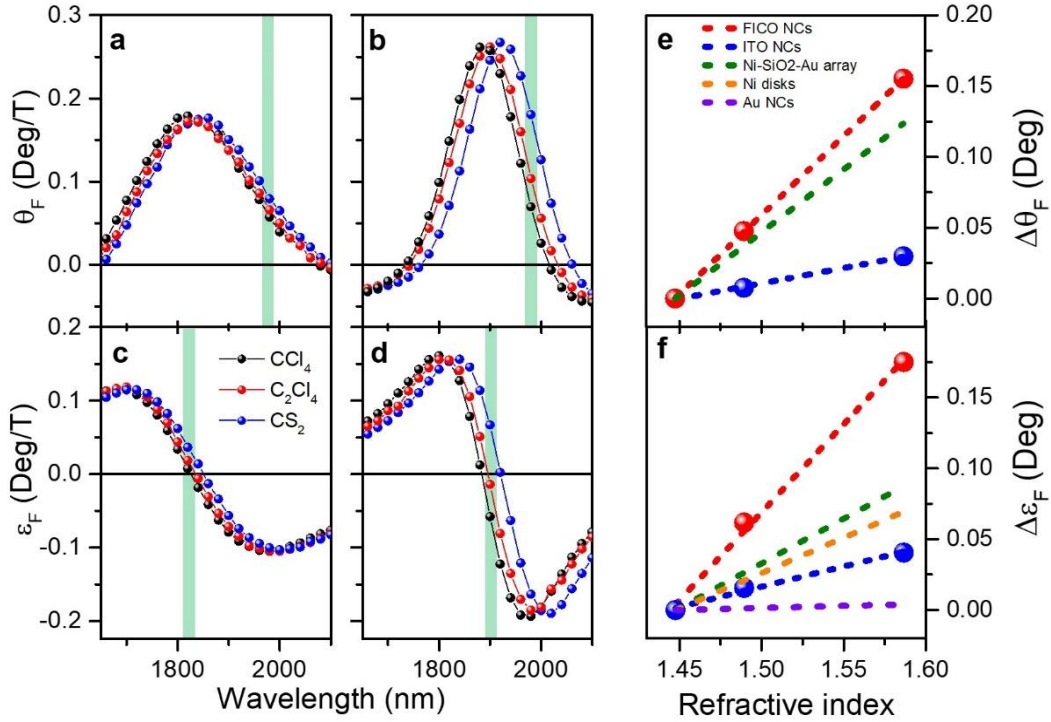

Figure 3: Refractometric sensing experiment. Normalized Faraday ellipticity (a,b) and rotation (c,d) spectra of ITO and FICO NCs dispersed in $CCl_4$ (black line), $C_2Cl_4$ (red line) and $CS_2$ (blue line). The variation of MO rotation and ellipticity with the refractive index at a fixed wavelength (highlighted in green in a-d) and for 1.4 Tesla of applied field is plotted in (e) and (f) respectively, where the dots represent the experimental points, while the dashed line is the linear fit. The values reported in the literature for Au spherical colloidal NCs,[25] Ni nanodisks[31] and Au/SiO2/Ni nanodisks array[53] are reported for comparison as dashed purple, orange and green lines respectively (details on the experimental conditions, i.e. applied field and wavelength range, are reported in Table S3).

In view of device applicability, a remark should be made on the magnetic field dependence of magnetoplasmonic effects. In ferromagnetic materials,[31,33] magnitude is proportional to magnetization. In nickel nanodisks, for instance, the magneto-optical signal saturates at ≈ 0.4 Tesla; a higher field value will not yield any signal increase. In non-magnetic TCOs, on the other hand, effects scale linearly with the applied field. This can be an advantage or a disadvantage (Ni can be saturated with a cheap NdFeB permanent magnet). However, with proper miniaturization, high magnetic field values can be achieved in portable devices (hard drive write heads apply fields up to 2.5 T), giving a net advantage to non-magnetic TCOs.

Considering that our setup is able to measure signals down to 0.33 mdeg ($1\times10^{-5}$ ΔA units, i.e. three times the standard deviation of the measurement), ΔRI down to $2.6\times10^{-4}$ can be easily detected. A magneto-optical set up specifically optimized for high sensitivity can detect signals down to $3\times10^{-6}$ mdeg ($1\times10^{-7}$ ΔA units),[54] affording RI sensitivities of $3\times10^{-6}$, close to the current technologies that employ the tracking of LSPR maximum measured in extinction spectroscopy.[55,56] A significant limitation of the latter approach lies in the fact that delicate curve fitting procedures are required to detect very small shifts of the wide LSPR resonance and the method can become unstable in a real analytical matrix. Our detection strategy, on the other hand, does not require a fitting procedure, but simply consists in measuring changes in intensity of the Faraday ellipticity at a fixed wavelength. In addition to the high sensitivity afforded by the sloping magneto-optical signal, our approach has the advantage of using an observable which is modulated both in polarization and magnetic field, thus making it very stable to matrix-related interference: virtually any interference (with the exception of magneto-optically active species) will be filtered out by the dual modulation.

In conclusion, ITO and FICO NCs despite being non-magnetic, display a significantly improved magneto-optical response, challenging state of the art ferromagnetic magnetoplasmonic nanomaterials.[31,33,53] This is afforded by the simultaneous presence of sharp plasmon resonances and low electron effective mass with respect to metals. The experimental spectra are in qualitative and quantitative agreement with our analytical model based on circular magnetoplasmonic modes. Proof of concept experiments demonstrate the applicability of the investigated NCs for magnetoplasmonic refractometric sensing, with dramatically improved performances compared to Au NCs and even superior with respect to the most promising magnetoplasmonic systems reported in the literature. Moreover, the sensitivity of our proposed approach is competitive with the current state of the art of refractometric sensing based on LSPR measured in extinction spectroscopy, with the advantage of not relying on a fitting procedure.

Considering the current growing interest in semiconductor NCs, in the near future the advancement in their synthesis and in the understanding of the correlation between structural parameters and the optical response could potentially lead to even sharper plasmonic resonances which could further increase the sensitivity of our proposed approach. Moreover, we believe that our understanding of the magneto-optical response in heavily-doped semiconductor NCs can trigger a new interest in these materials for magnetoplasmonics. To this aim, the use of heavily doped semiconductor NCs as building blocks for hybrid nanostructures combining them with a magnetic unit[57] or co-doping them with magnetic ions,[41,58–60] would represent an interesting future perspective, which has not yet been exploited for applications in magnetoplasmonics.

**Methods**

**Synthesis of ITO NCs**. 1.8 mmol of In(acac)$_3$ and 0.2 mmol of Sn(acac)$_2$Cl$_2$ were dissolved in 20 mL of 1-octadecene with 6 mmol of oleylamine and 6 mmol of oleic acid. Under constant magnetic stirring and under vacuum the reaction mixture was heated to 80 °C over 10 min and maintained for 30 min. Then, under a nitrogen atmosphere, the temperature was increased to 160 °C over 12 min and maintained for 1 h, after which it was increased to 310°C over 26 min and maintained for additional 2 h. The mixture was cooled down to room temperature by removing the heat source. The NCs were washed twice by centrifugation in ethanol at 4400 g for 5 minutes, after which they were dispersed in hexane.

**Synthesis of FICO NCs**. The synthetic procedure was adapted from Ye et al.[45] 1.2 mmol of Cd(acac)$_2$ and 0.3 mmol of InF$_3$ were added to 5.25 mmol of oleic acid and 50 mL of 1-octadecene. The mixture was heated to 120 °C for 1 h under vacuum, after which the temperature was raised to reflux (320 °C) during 18 min under Ar atmosphere. After 15 min at reflux the mixture changed colour to dark green, and after additional 10 min the reaction was stopped by removing the heating source and the mixture was cooled down to room temperature.

The NCs were washed twice by centrifugation in isopropanol at 4400 g and dispersion in hexane. Bulky secondary products were removed by centrifugation in hexane at 1000 g and discarding the bulky precipitate.

**XRD**. Powder X-ray Diffraction (XRD) measurements were carried out using a Bruker D8 Advance diffractometer equipped with a Cu K$_\alpha$ radiation and operating in Theta-Theta Bragg Brentano geometry at 40 kV and 40 mA.

**TEM**. Transmission Electron Microscopy (TEM) analysis was performed using a JEOL 100 SX, operating at 100 kV. Samples were prepared by drop drying a dilute suspension of NCs in hexane onto 200 mesh carbon-coated copper grids. The mean size and size distribution were obtained from statistical analysis over at least 300 NCs.

**ICP-AES**. ICP-AES measurements were performed in triplicate by using a Varian 720-ES Inductively Coupled Plasma Atomic Emission Spectrometer, on samples (≈1 mg) digested in concentrated aqua regia (HCl suprapure and HNO$_3$ sub-boiled in 3:1 ratio) and in the presence of H$_2$O$_2$, diluted with ultrapure water (≥18 MΩ) and analyzed using Ge as internal standard. Calibration standards were prepared by gravimetric serial dilution from monostandard at 1000 mg/L. The wavelengths used for In, Sn and Cd were 325.6, 189.9 nm and 214.4 nm respectively. The dopant content is defined as $\frac{n_D}{n_D+n_C} \cdot 100$, where $n_D$ and $n_C$ are the moles of substitutional dopant and lattice cation respectively.

**UV-vis-NIR Extinction measurements**. Extinction spectra were collected using a commercial Cary5000 UV-vis-NIR spectrophotometer (Agilent), using a 1 mm quartz QX cuvettes. Solvents transparent in the NIR (CCl$_4$, C$_2$Cl$_4$ and CS$_2$) were used to disperse the NCs for optical and magneto-optical characterization.

**Magneto-optical characterization**. Magneto-optical characterization was performed in the Faraday configuration using in house built MO set up in the spectral range 800-2500 nm, and exploiting the polarization modulation technique using a photo-elastic modulator (PEM) which modulates the light polarization alternatively between RCP and LCP at a frequency of 47 kHz. Light provided by the Xe lamp (power 300 W) passes through a monochromator (Oriel Cornerstone 260) and then is linearly polarized at 90° using a Rochon polarizer. Linearly polarized light is then focused on the sample which is placed in an electromagnet able to apply a magnetic field of 1.4 Tesla. The polarization of light transmitted by the sample is then modulated by the PEM, whose retardation is set to 0.383λ, after which an analyser (Glan-Thompson prism) oriented at 45° is placed before an InGaAs detector. The output from the detector is analysed by two lock-in amplifiers, locked respectively to the first and the second harmonic of the PEM, collecting respectively the signal related to Magnetic Circular Dichroism or Faraday ellipticity ($\varepsilon_F$) and Faraday rotation ($\theta_F$). To filter out the residual environmental light, the signal is further modulated at 440 Hz by a mechanical chopper. A third lock-in amplifier locked to the chopper is used to retrieve the total light collected by the detector. The MO signal is calibrated using a $K_3Fe(CN)_6$ solution as a standard reference sample,[61] in order to calibrate the ratio between the AC and the DC signal that reaches the detector. A Hall probe is employed to measure the effective magnetic field in the sample position. The spurious dichroic signal of the set up was subtracted by measuring the MO signals at +1.4 Tesla and -1.4 Tesla: the semi-difference between the two gives the MO signal, assuming that the natural dichroism and optical rotation are invariant with the magnetic field. The measurements were performed in solution, using 1 mm quartz QX cuvettes. The Faraday rotation and ellipticity of the solvent were collected in the same experimental conditions and subtracted from the signal of the sample, in order to isolate the contribution of the NCs. The MO signal was then normalized for the optical density and for the applied magnetic field, and converted into ellipticity and rotation angles (in degrees) according to previous work.[52,54] The optical density of the NC dispersions analysed was in the range 0.7-1.4, which is in the optimal range that maximizes the signal-to-noise ratio.[52] The MO spectra reported in the main text are collected at 1.4 Tesla, and normalized for the optical density and the applied magnetic field.

**Analytical Calculation of the optical and magneto-optical response.** The calculation of the optical and magneto-optical response was performed with an analytical model previously developed.[25,51] The Drude dielectric function of the semiconductors was inserted in the quasi-static field- and helicity-dependent polarizability (equation S 4). The fitting of extinction and ellipticity spectra was performed using the same analytical model, extracting carrier parameters $N$, $m$ and $\gamma$. More details are provided in section S2 of the supporting information.

**Acknowledgment**

Authors acknowledge the financial support of H2020-FETOPEN-2016-2017 Grant No. 737709 FEMTOTERABYTE (EC) and of PRA_2017_25 (Università di Pisa). Dr. Mirko Severi is acknowledged for ICP-AES characterization. Dr. Paolo Lucchesi is acknowledged for the assistance in TEM measurements.

**Author Contributions**

F.P. and A.N. proposed the concept. A.G. and F.P. designed the experimental magneto-optical set up. A.G. synthesized ITO and FICO NCs, performed the measurements, analysed the data and performed the analytical calculations. A.G. and F.P. wrote the paper, with input and contribution from M.G., C.S., B.T. and A.N.

**Corresponding Authors**

*francesco.pineider@unipi.it

**Supporting Information.** Size distribution of the nanocrystals obtained from the analysis of TEM images. Further details on the analytical model and fitting procedures employed. Refractive index of the solvents employed in the experiments as a function of the wavelength. Extinction spectra of ITO and FICO NCs dispersed in $CCl_4$, $C_2Cl_4$ and $CS_2$. Comparison of the magnetoplasmonic performances of FICO and ITO NCs with the most promising magnetoplasmonic systems reported in the literature.

# Supporting Information

## S1. Structural characterization

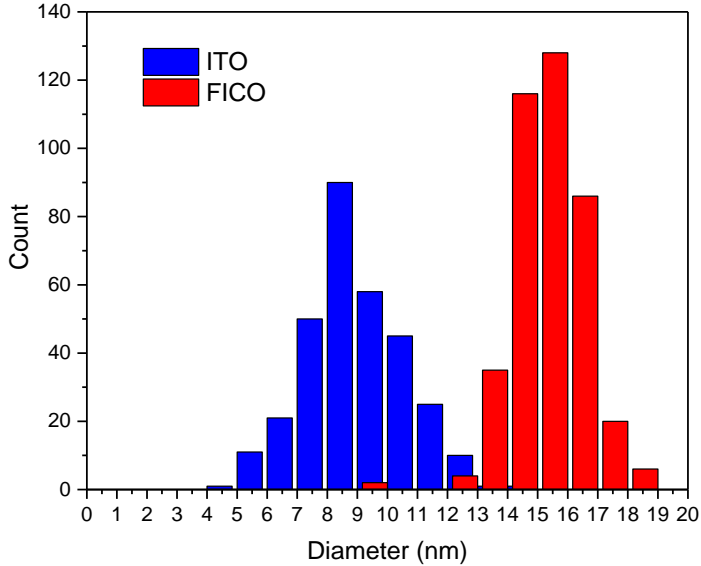

Figure S 1: Size distribution of ITO and FICO NCs, determined from TEM statistics.

## S2. Analitical model and fitting of extinction and magneto-optical spectra

To fit the normalized extinction and MCD spectra, a matlab routine was developed, according to an analytical model introduced in previous work.[1,2]

In the dipolar quasi-static approximation, the polarizability of a spherical NC of size D and dielectric function $\varepsilon(\omega)$ in a dielectric medium with dielectric constant $\varepsilon_m$, can be expressed through Equation S 1, and can be related to the extinction cross section (Equation S 2):[3]

Equation S 1    $\alpha(\omega) = -\frac{\pi D^3}{2} \frac{(\varepsilon(\omega) - \varepsilon_m)}{(\varepsilon(\omega) + 2\varepsilon_m)}$

Equation S 2    $\sigma(\omega) = k\sqrt{\varepsilon_m} Im[\alpha(\omega)]$

MO effects in purely plasmonic NCs are rationalized in terms of the Lorentz force acting on free electrons:[2,4,5]

Equation S 3    $m\frac{dv}{dt} + \gamma m v = -eE - ev \times B$

where *e* and *m* are the charge and effective mass of the electron, *v* its velocity, $\gamma$ is the damping parameter and *B* is the external magnetic induction. Commonly, the effect of the magnetic field on the dielectric function of a metal is treated adding an off-diagonal term in the dielectric tensor or in the polarizability tensor, which takes into account for the magnetic-field driven modification of the dielectric function.[4,6] Alternatively, the problem can be also solved by using a diagonal form of the polarizability, as reported by Gu and Kornev,[1] who, exploiting the fact that the term $ev \times B$ is smaller with respect to the other term, solved Equation S 3 perturbatively. Following the formulation by Gu and Kornev, the field- and helicity-dependent polarizability can be written in a diagonal form according to Equation S 4**Errore. L'origine riferimento non è stata trovata.**:[1,2]

Equation S 4  $\alpha_B(\omega) = -\dfrac{\pi D^3}{2} \dfrac{(\varepsilon(\omega)-\varepsilon_m)+B(f(\omega)-f_m)}{(\varepsilon(\omega)+2\varepsilon_m)+B(f(\omega)-f_m)}$

where the second term at the numerator and denominator describes the effect of the magnetic field on the polarizability, in which $f(\omega) = f_1 + if_2$ and $f_m$ are the coupling functions describing the interaction with the magnetic field for the NC and the medium. At zero applied magnetic field, Equation S 4 is simplified to the well-known quasi-static polarizability of a sphere (Equation S 1).

Considering the symmetry of the problem, a change in helicity is equivalent to an inversion of the direction of the external magnetic field B. It follows that the differential polarizability (Equation 2 in the main text) can be obtained from the difference between the polarizability calculated with positive and negative applied field. Using the obtained $\Delta\alpha_B$, the helicity-dependent normalized differential cross section can be readily obtained through Equation S 5.

Equation S 5  $\Delta\sigma_{norm}(\omega) = \dfrac{\Delta\sigma(\omega)}{\sigma_{max}} = \dfrac{k\sqrt{\varepsilon_m} Im[\Delta\alpha_B(\omega)]}{\sigma_{max}}$

The $\Delta\sigma$ can be converted from differential absorption units into ellipticity angle ($\varepsilon_F$) using Equation S 7. The following equations were thus employed to fit the normalized extinction and ellipticity spectra:

Equation S 6  $\sigma_{norm}(\omega) = \dfrac{\sigma(\omega)}{\sigma_{max}} = \dfrac{k\sqrt{\varepsilon_m} Im[\alpha(\omega)]}{\sigma_{max}}$

Equation S 7  $\varepsilon_F \, (deg) = ln10 \cdot \dfrac{\Delta\sigma_{norm}(\omega)}{4} \cdot \dfrac{180}{\pi}$

where $\alpha_B$ and $\Delta\alpha_B$ are calculated according to Equation S 4 and Equation 4 (main text), for an applied field of 1 Tesla. The Drude dielectric function is used for the NCs:[3]

Equation S 8  $\varepsilon(\omega) = \varepsilon_\infty - i\dfrac{\omega_p^2}{\omega}\dfrac{\gamma - i\omega}{\omega^2 + \gamma^2} = \varepsilon_\infty - i\dfrac{e^2 N}{m\varepsilon_0 \omega}\dfrac{\gamma - i\omega}{\omega^2 + \gamma^2}$

where $\varepsilon_\infty$ is the background permittivity of the semiconductor NC, $\gamma$ is the electron damping parameter, $N$ and $m^*$ are the free electron density and effective mass, $\varepsilon_0$ is the vacuum permittivity, $e$ is the electron charge, $\hbar$ the barred plank constant, and $\omega_P = \sqrt{\dfrac{e^2 N}{m\varepsilon_0}}$ is the plasma frequency.

For the helicity-dependent polarizability, Equation S 9 is used for the coupling function of the oxide, following the Drude formulation reported by Gu and Kornev,[1,2] while a $f_m$ of $1.06 \cdot 10^{-6}$ is used for the solvent.[2]

Equation S 9 $\quad f(\omega) = -\frac{e^3 N}{m^2 \varepsilon_0 \omega} \frac{(\gamma - i\omega)^2}{(\gamma^2 + \omega^2)^2}$

Equation S 8 and Equation S 9 are inserted into Equation S 4 and then into Equation S 6 and Equation S 5, to obtain the fitting functions. Using these equations, the normalized extinction and ellipticity spectra were fitted (Figure S 2 and Figure S 3) simultaneously, using *N*, *m* and *γ* as common parameters for the two fitting functions. $\varepsilon_m$ of the solvent (see section S3) was used in Equation S 4, while values of $\varepsilon_\infty$ of 4.0 and 5.6 were used respectively for ITO and FICO NCs, taken from the literature.[7,8] *N*, *m\** and *γ* were obtained from the fitting. For ITO NCs a frequency dependent damping function ($\gamma(\omega)$, Equation S 10) was used to account for the strong electron scattering caused by ion impurities,[7,9] while for FICO NCs, a constant value of *γ* was sufficient to fit the spectra.

Equation S 10 $\quad \gamma(\omega) = \gamma_L - \frac{\gamma_L - \gamma_H}{\pi} [\tan^{-1}\left(\frac{\omega - \omega_X}{\omega_W}\right) + \frac{\pi}{2}]$

Indeed, dopant impurities are known to give rise to electron-scattering which is generally rationalized with the presence of two damping regimes: one at high frequency, characterized by a damping constant $\gamma_H$, where the electrons can escape the Coulomb interaction with the impurity ions, as they travel faster; and one at low frequency, characterized by $\gamma_L$, where the damping is higher.[9,10] The two regimes are separated by a frequency threshold $\omega_X$, while the width of the crossing region between the two regimes is expressed by $\omega_W$. The different behaviour between ITO and FICO can be explained with the fact that $Sn^{4+}$ ions scatter electrons more efficiently than $In^{3+}$,[11] due to the higher charge. In addition, the electron scattering is negligible for the F- co-dopant in FICO, due to its lower charge with respect to the $Sn^{4+}$ and $In^{3+}$ impurities.[12] Alternative fitting approaches reported in the literature reproduced with good agreement the asymmetric line shape of the LSPR in ITO NCs by considering a distribution of carrier densities.[7] Employing the latter model similar agreement with the experimental extinction and MCD was obtained (not reported here), as well as comparable free carrier parameters.

Parameters obtained from the fitting are reported in Table S1, while the $\gamma(\omega)$, used for ITO is reported in Figure S 3 C and Table S 2.

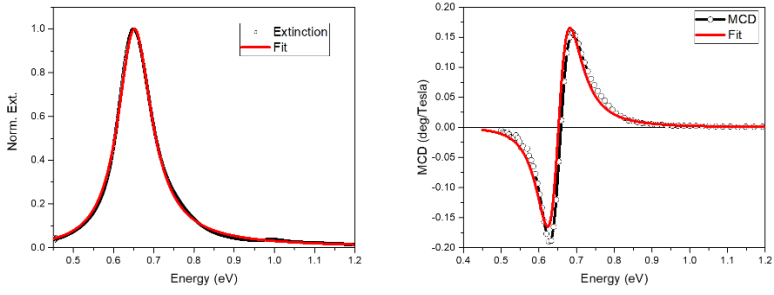

Figure S 2: Fitting of normalized extinction and MCD spectrum for FICO NCs.

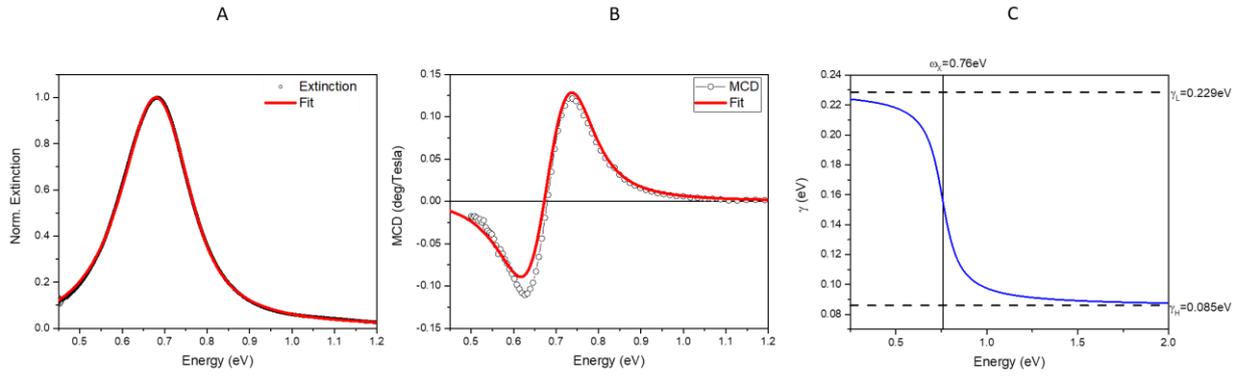

Figure S 3 Fitting of normalized extinction (A) and MCD spectrum (B) for ITO NCs; C) frequency dependent damping function extracted from the fitting.

Table S 1: LSPR parameters and Drude parameters obtained from the analysis of the extinction and ellipticity spectra.

|      | $\lambda_{LSPR}$ (nm) [a] | FWHM (eV) [a] | $N$ (cm$^{-3}$) [b] | $m$ [b] | $\omega_c$ (meV/Tesla) [c] | $\omega_c/\gamma \times 1000$ (T$^{-1}$) [d] |
|------|---------------------------|---------------|---------------------|---------|---------------------------|----------------------------------------------|
| Au   | 520                       | 0.410         | 5.57 x $10^{22}$    | 0.99$m_e$ | 0.114                   | 0.28                                         |
| ITO  | 1825                      | 0.190         | 7.13 x $10^{20}$    | 0.27$m_e$ | 0.428                   | 2.25                                         |
| FICO | 1883                      | 0.103         | 9.24 x $10^{20}$    | 0.31$m_e$ | 0.372                   | 3.72                                         |

a) obtained from the analysis of the extinction spectrum; b) obtained from the simultaneous fit of extinction and ellipticity spectra. Data for Au are taken from references;[20,39] c) cyclotron frequency calculated from electron effective mass using the following relation: $\omega_c = eB/m$, calculated for 1 Tesla of applied field. d) magnetoplasmonic quality factor calculated as $\omega_c/\gamma$.

Table S 2: Damping parameters of obtained from the fit of extinction and MCD spectra for ITO and FICO NCs.

|      | $\gamma_L$ (eV) | $\gamma_H$ (eV) | $\omega_x$ (eV) | $\omega_w$ (eV) |
|------|-----------------|-----------------|-----------------|-----------------|
| ITO  | 0.229           | 0.085           | 0.76            | 0.068           |
| FICO | 0.102           | -               | -               | -               |

After obtaining the Drude parameters from the simultaneous fit of extinction and MCD spectra, these parameters are used to calculate the Faraday rotation (employing Equation S 11), which is reported in Figure 2 e,f (main text).

Equation S 11 $\quad \vartheta_{F\,norm} = \frac{\vartheta_F}{\sigma_{max}} = \frac{k\sqrt{\varepsilon_m}Re[\Delta\alpha_B]}{\sigma_{max}}$

With an alternative approach, a simultaneous fit of extinction, ellipticity and rotation spectra using Equation S 6, Equation S 5 and Equation S 11 can be performed with excellent results, as shown in Figure S 4 for FICO NCs, obtaining free electron parameters comparable to those obtained with the previously discussed approach. However, we pointed out that the measurement of Faraday rotation can be challenging especially for the subtraction of the strong solvent contribution, and in this work we employed the approach based on fitting of ellipticity and extinction spectra to extract Drude parameters as it is more easily addressable experimentally.

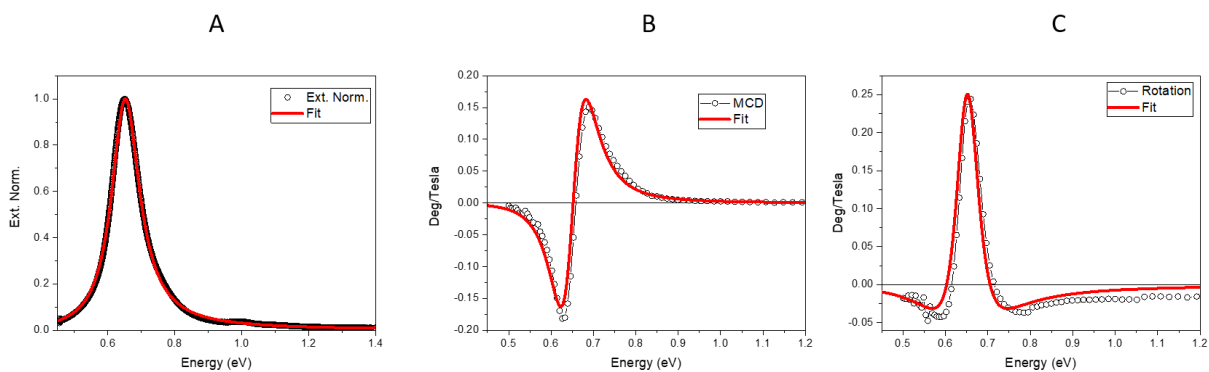

Figure S 4: Simultaneous fitting of extinction (A), MCD (B) and Faraday rotation (C) spectra for FICO NCs dispersed in CCl$_4$.

## S3. Sensing experiments

The refractive index of the solvents used in the main text were taken from the literature.[13,14] As can be seen in Figure S 5, the dispersion curve is quite flat in the region of interest (1200-2400 nm) for all the three solvents, which makes reasonable for us to take a constant value for the refractive index of the solvent. Values of 1.4477, 1.4895 and 1.5866 are used for $CCl_4$, $C_2Cl_4$ and $CS_2$, respectively.

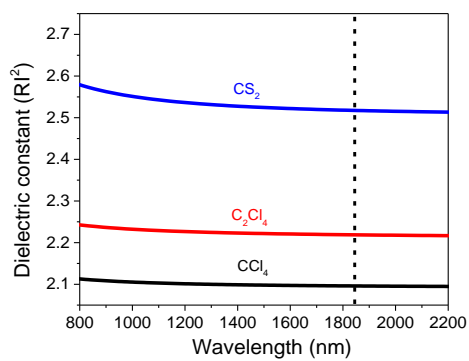

Figure S 5: Dielectric constant vs wavelength curve for the solvent used in this work. Data from Kedenburg et al.[13] are used for $CCl_4$ and $CS_2$, while data from Chemnitz et al.[14] are used for $C_2Cl_4$. The dashed vertical line indicates the region of the LSPR peak for our samples.

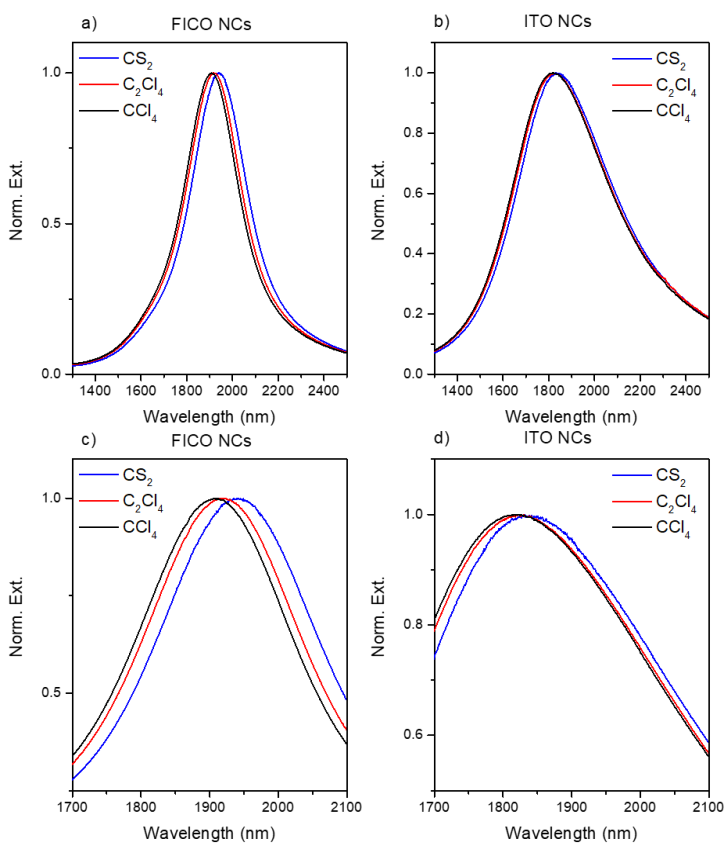

Figure S 6: Normalized extinction spectra of FICO (a) and ITO (b) NCs dispersed in three different solvents with different refractive index and magnification in the region of the LSPR maximum c, d).

**S4. Comparison of magnetoplasmonic performances**

Table S 3 Comparison of magnetoplasmonic performances of ITO and FICO NCs with the most promising systems reported in the literature.

|  | Au NPs [i] | Ni [ii] | Ni array [iii] | Au/SiO2/Ni array [iii] | ITO NCs [iv] | FICO NCs [iv] |
|---|---|---|---|---|---|---|
| [a] $\lambda_{LSPR}$ (nm) | 520 | 650 | 800 | 870 | 1825 | 1880 |
| [b] $\Delta\lambda/\Delta RI$ (nm/RIU) | 80 | 250 | 214 | 211 | 184 | 223 |
| [c] $\Delta\varepsilon/\Delta RI$ (deg/RIU) | 0.029 | 0.50 | 0.44 | 0.63 | 0.29 | 1.24 |
| [d] $\Delta\theta/\Delta RI$ (deg/RIU) | N.A. | N.A. | 0.64 | 0.89 | 0.22 | 1.12 |
| [e] B (Tesla) | 1.3 | 0.4 | 0.4 | 0.4 | 1.4 | 1.4 |

*a) position of the LSPR peak in extinction; b) refractive index sensitivity of the LSPR extinction peak; c,d) refractive index sensitivity of the magneto-optical signal (ellipticity and rotation); e) Applied field employed in the experiments. (i) Data from Reference 2. (ii) Data from reference 15. (iii) Data from reference 16. (iv) Data from the present work.*